\documentstyle[epsf,rotate]{elsart}

\newcommand{\figdir}{./}

\begin{document}

\begin{frontmatter}

\title{Energy avalanches in a rice-pile model}

\author[IFF]{Lu\'{\i}s A. Nunes Amaral\thanksref{LANA}}
\author[NBI]{Kent B{\ae}kgaard Lauritsen\thanksref{KBL}}

\thanks[LANA]{L.Amaral@kfa-juelich.de}
\thanks[KBL]{baekgard@nbi.dk}

\address[IFF]{Institut f\"ur Festk\"orperforschung, 
        Forschungszentrum J\"ulich \\ D-52425 J\"ulich, Germany}
\address[NBI]{Niels Bohr Institute, Center for Chaos and Turbulence Studies \\ 
        Blegdamsvej 17, 2100 Copenhagen \O, Denmark}

\date{March 28, 1996}


\begin{abstract}
  We investigate a one-dimensional rice-pile model.  We show that the
  distribution of dissipated potential energy decays as a power law
  with an exponent $\alpha=1.53$.  The system thus provides a
  one-dimensional example of self-organized criticality.  Different
  driving conditions are examined in order to allow for comparisons
  with experiments.
\end{abstract}


\end{frontmatter}


\section{Introduction}

Statistical-mechanical investigations of driven nonequilibrium systems
is a field of much current interest.  A special class of such systems
are those which reach the steady state through a self-organizing
dynamics consisting of avalanches propagating through the system. If,
in addition, the steady state is characterized by a power-law
distribution for the sizes of avalanches the behavior is referred to
as self-organized critical (SOC) \cite{BTW}.  Simple ``sandpile''
models have been introduced in order to illustrate the SOC behavior
\cite{BTW,KNWZ,CL,FF,Toner,Frette,HK,luebeck}.  Real sandpiles,
however, have turned out to display noncritical behavior---i.e.,
they are described by avalanche size distributions with a characteristic
scale \cite{JLN,H,RVK,RVR,Feder,Hernan}.

Recently, an experiment with rice grains was performed
\cite{frette-etal:1996}. It was found that a pile of elongated rice
grains evolved into a SOC state with the distribution of dissipated
potential energy $E$ being flat for small avalanches and crossing over
to a power law with exponent $\alpha_{\rm exp} \simeq 2.0$ for large
avalanches.  Physically motivated rice-pile models have been
introduced which indeed do display SOC behavior in one dimension
\cite{amaral-lauritsen:1996a,christensen-etal:1996,
  markosova-etal:1996,amaral-lauritsen:1996b} but with an exponent
which differs from that observed experimentally.

In this paper, we measure the energy avalanches for the rice-pile
model which we introduced in Ref.\ \cite{amaral-lauritsen:1996a}.
Previously we found that the distribution of avalanches defined as
the number of topplings $s$ followed the power-law form
$P(s, L) \sim s^{-\tau} \, f_s(s/L^{\nu_s})$,
with $\tau = 1.53$ and $\nu_s=2.20$ \cite{amaral-lauritsen:1996a}. 
Here, the distribution $P(E, L)$ of energy avalanches in a system of size
$L$ is measured.  We find that $P(E, L)$ scales with the same exponent
as $P(s, L)$ in accordance with the fact that a toppling event on
average dissipates a constant energy.  We change the drive in our model
in order to make comparisons with the experimental results in
\cite{frette-etal:1996} more quantitative. Despite the fact that this
leads to a flat distribution for small $E$, in perfect agreement with
the experiment, the power-law tail for large $E$ is still described by an
exponent 1.53.

\section{Rice-pile model}

We briefly describe the model we introduced in
\cite{amaral-lauritsen:1996a}.  We consider a one-dimensional system
of length $L$ with a wall at $i=0$ and an open boundary at $i = L+1$,
where particles fall off the pile. The dynamics of the model consists
of deposition and relaxation: At each time step one grain is added at
$i=1$.  Then, the pile is allowed to relax in order to reach a new
stable configuration.  The relaxation process is considered to be fast
compared to the deposition time scale. During the relaxation {\it
  active\/} columns topple one grain from $i$ to $i+1$ with
probability $p({\delta h}_i)$, where ${\delta h}_i \equiv h(i) -
h(i+1)$ is the local slope.  A column $i$ of the pile is said to be
active if, in the anterior time step, it (i) received a grain from
column $i-1$, (ii) toppled a grain to column $i+1$, or (iii) column
$i+1$ toppled one grain to its right neighbor.  The probability
$p({\delta h}_i)$ to move a grain is taken to be:
\begin{equation}
        p({\delta h}_i) = \left\{
                \begin{array}{ll}
                        0,  &       {\delta h}_i \le S_1,  \\
                        p,  & S_1 < {\delta h}_i \le S_2,  \\
                        1,  & S_2 < {\delta h}_i ,
                \end{array}
                \right.
                                        \label{eq:p(deltah_i)}
\end{equation}
where we will use $S_1=1$, $S_2=4$, and $p=0.6$; the results
are insensitive to the precise values of these parameters.

Physically the parameter $p$ describes the friction between rice
grains.  It also incorporates the possibility that a metastable
packing configuration will be reached when a grain topples.  The
friction effect is the new ingredient in the model compared to other
models and it introduces a large range of slopes in the rice pile
instead of a single critical value.  The parameter $S_1$ accounts for
the fact that small slopes are stable.  The parameter $S_2$ models the
effect of gravity on the packing configurations.  We assume that above
the maximum value $S_2$ of the local slope, it is no longer possible
for a local stable configuration to be achieved, thus a grain must be
toppled. In the limiting cases $p=0, 1$, or $S_2=S_1$, we recover the
model in Ref.\ \cite{BTW} (which has trivial behavior in one
dimension).

\section{Avalanche results}

\begin{figure}
\centerline{
\epsfysize=0.85\columnwidth{\rotate[r]{\epsfbox{\figdir/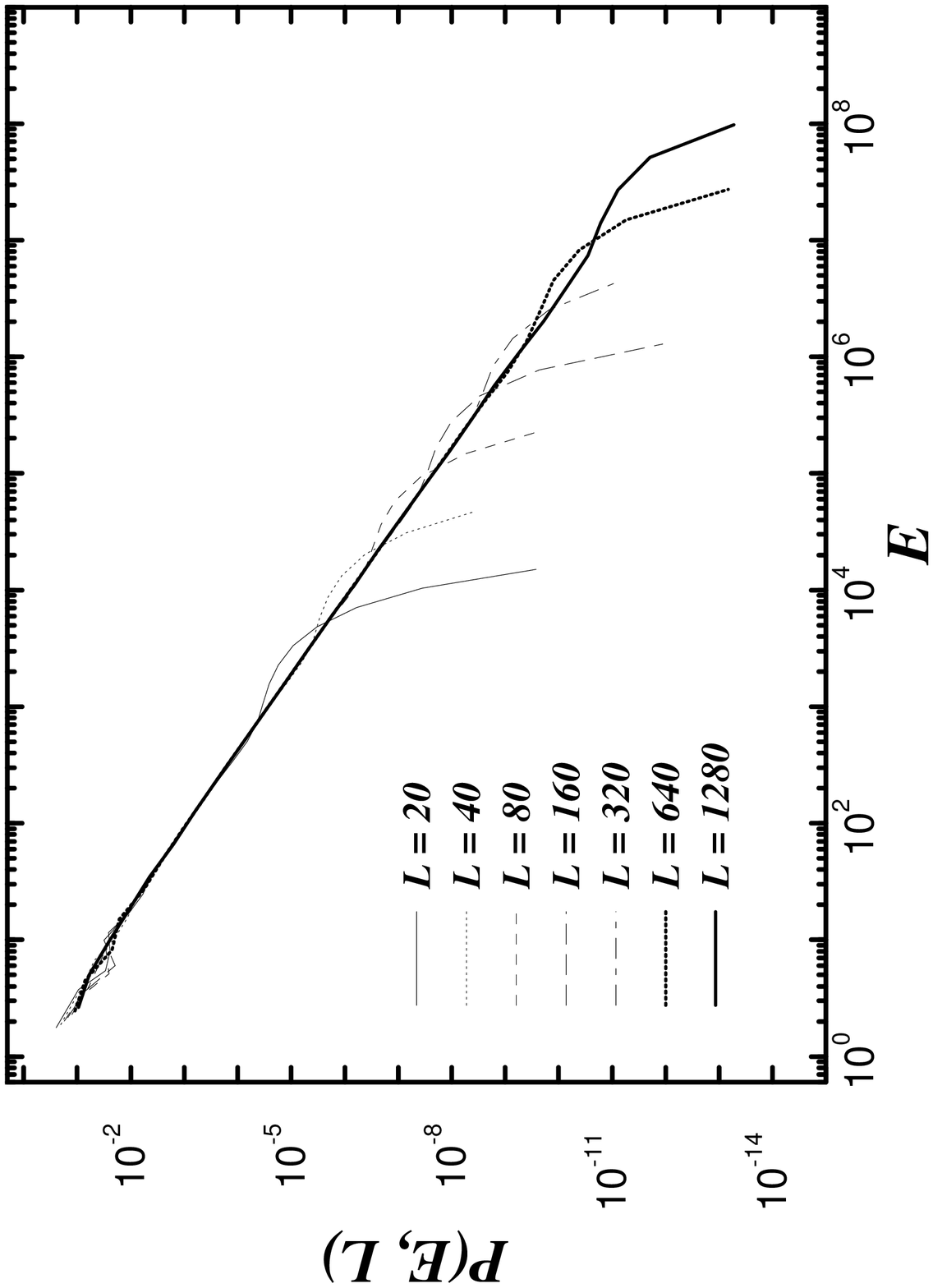}}}}
\vspace*{0.5cm}
\centerline{
\epsfysize=0.85\columnwidth{\rotate[r]{\epsfbox{\figdir/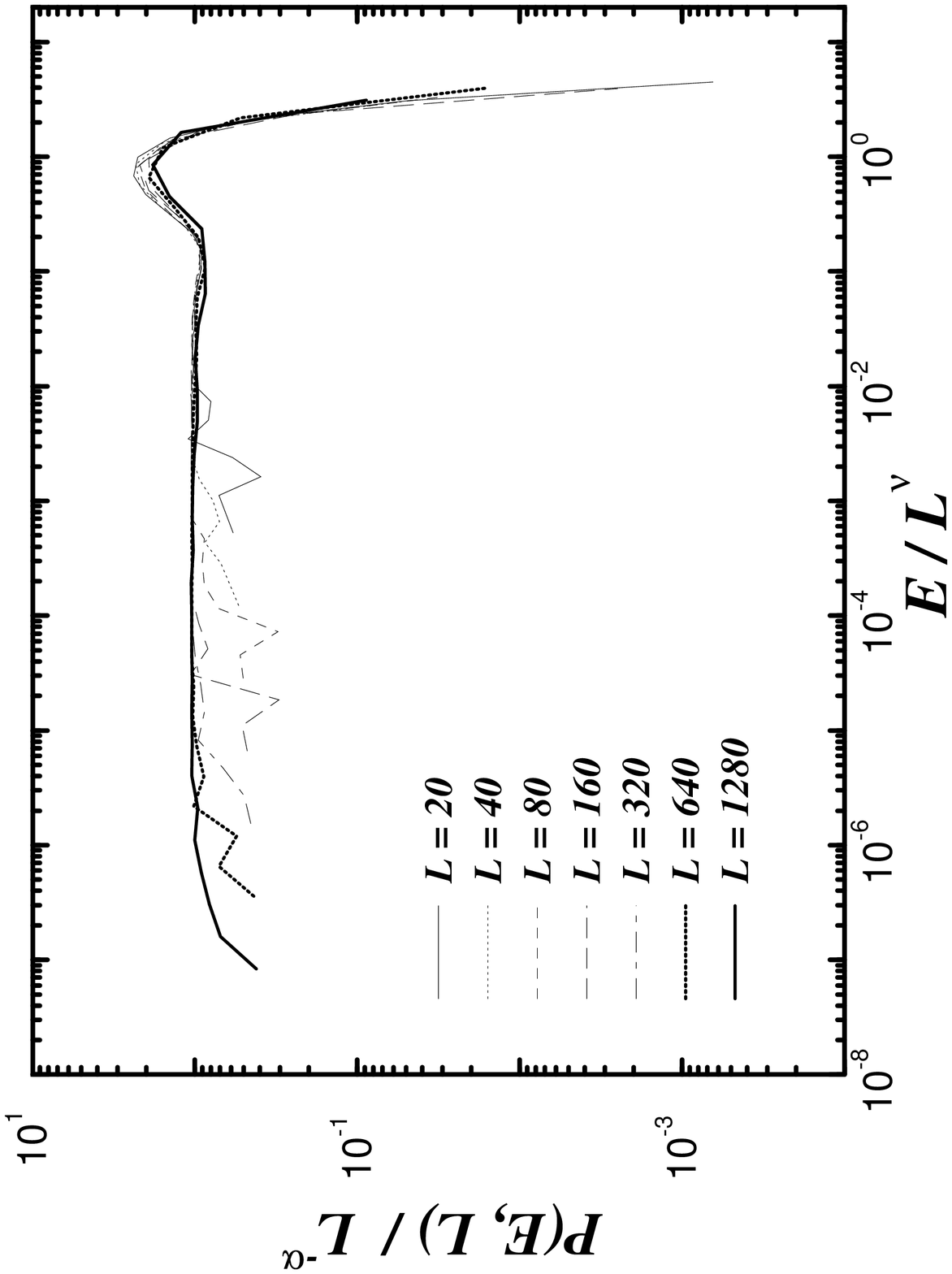}}}}
\vspace*{0.5cm}
\caption{ (a) Log-log plot of the probability density $P(E, L)$ for
different values of $L$.  The results was obtained with the parameters
$p=0.6$, $S_1 = 1$, and $S_2 = 4$. The data follows a power law
distribution for several decades with a cutoff that depends on the
system size.  (b) Data collapse of the curves displayed in (a)
according to Eq.~(\protect\ref{e-slow}) with the exponents $\alpha
\simeq 1.53$ and $\nu \simeq 2.20$.}
\label{f-slow}
\end{figure}

We study the model in the slowly driven limit where the rate of
deposition is slow enough that any avalanche, that might be started by
a deposited grain, will have ended before a new grain is deposited.
The simulations of the model show the existence of a SOC steady state.
We follow the definition of Ref.\ \cite{frette-etal:1996} and
calculate the size of avalanches as the dissipated potential energy in
between snapshots of the profile.  Figure~\ref{f-slow}(a) shows the
probability density of avalanche sizes for different system sizes. We
find that the scaling form
\begin{equation}
        P(E, L) \sim E^{-\alpha}~f(E / L^{\nu})
\label{e-slow}
\end{equation}
describes the distribution of avalanche sizes.  The validity of
Eq.~(\ref{e-slow}) is reassured by the good data collapse displayed in
Fig.~\ref{f-slow}(b), where we used the values $\alpha = 1.53 \pm
0.05$ and $\nu = 2.20 \pm 0.05$. 
Since a toppling on average dissipates a constant energy,
we expect that $\alpha=\tau$ (and $\nu = \nu_s$) as
observed numerically.
By assuming that the average value is $\left< E \right> \sim L$
in the critical state,
it follows from Eq.~(\ref{e-slow}) that $\alpha=2-1/\nu$
\cite{maya}. This relation is in nice agreement with our numerical values
for the exponents.
We note that the scaling function
$f$ has a peak for values of the argument close to the cutoff region.
This is a finite-size effect which is due to the possibility
to form a supercritical state which then relaxes
through a very large avalanche \cite{amaral-lauritsen:1996a}.

Our estimate for the exponent $\alpha$ is different from the value
$\alpha_{\rm exp} \simeq 2$ reported in \cite{frette-etal:1996}, and
the scaling function is also quite different from that observed
experimentally.  Part of the reason for this disagreement is due to
the fact that the experimental conditions were different from how we
measure the avalanche distribution $P(E, L)$.  The next step is thus
to study the model under conditions as similar as possible to the
experimental ones.  In the experiment grains were deposited randomly
in time with a rate of 4-6 grains between 
successive snapshots of the pile taken every 15 seconds
\cite{frette-etal:1996}.  Furthermore, since the profile of the pile
was not known at every time instant when grains were deposited, the
potential energy of the deposited (4-6) grains was estimated from the
last profile obtained.  The uncertainty in the added potential energy
induces a noise level proportional to the system size (see below).

To study the model under the ``experimental conditions'' we made the
following assumptions: (i) snapshots of the profile of the pile are
taken every $N_s$ time steps, and (ii) new grains are deposited on the
pile at an average rate of $1/N_d$.  Thus, every time step there is a
probability $1/N_d$ of a new grain being deposited.  In our
simulations, characteristic values for $N_s$ and $N_d$ were $10^4$ and
$2100$, respectively.

\begin{figure}[htb]
\centerline{
    \epsfysize=0.85\columnwidth{\rotate[r]{\epsfbox{\figdir/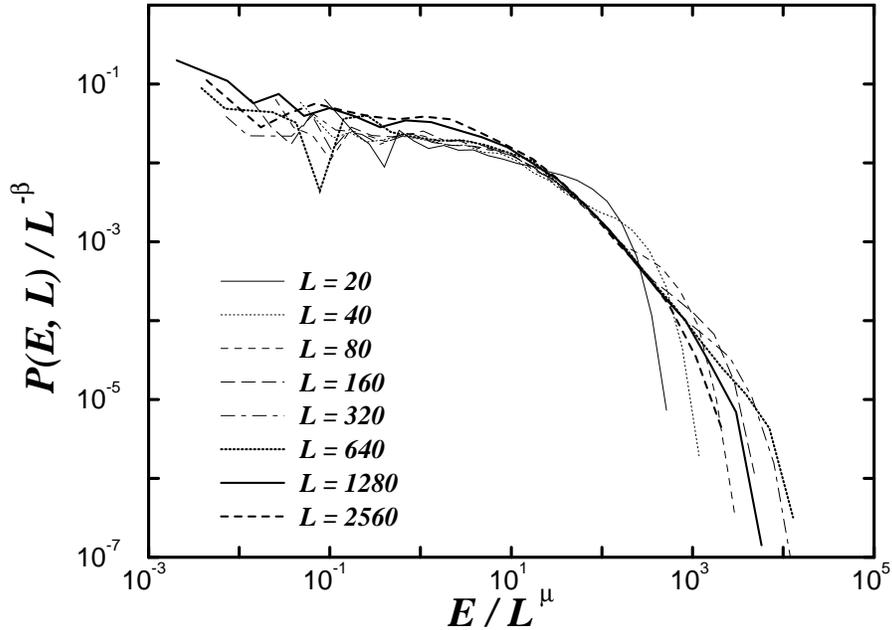}}}
}
\vspace*{0.5cm}
\caption{ (a) Log-log plot of the probability density of avalanche
sizes for different systems sizes.
The data was obtained for conditions similar to the experimental
ones but with an accurate determination of the dissipated energy. 
A power law dependence is observed with an exponent around 1.5. 
(b) Data from the same run as in (a) but now calculating the
dissipated energy according to the method used in the experiment.  It
is remarkable how different the data looks from (a).  A plateau for
small $E$ is visible and a dependence on $L$ is detectable. 
(c) Data collapse of the curves shown in (b) according to
Eq.~(\protect\ref{e-fdrv}) with the exponents $\beta = \mu \simeq 1$.
}
\label{f-fdrv}
\end{figure}

Figure~\ref{f-fdrv}(a) shows the probability density of avalanche
sizes for an accurate calculation of $E$.  No flat part is observed
for small values of $E$ and no dependence on $L$ is detected---except
for finite-size effects.  However, when we estimate $E$ as it was done
in the experiment of Ref.~\cite{frette-etal:1996} a significant
change occurs.  As shown is Fig.~\ref{f-fdrv}(b), a plateau whose
height depends on $L$, and originates from the uncertainty in the added
potential energy, is observed for small $E$. For large $E$, there is a
crossover to a power-law behavior at a value that scales with $L$.  In
fact, the data resemble quite well the experimental results.  We find
that the data in Fig.~\ref{f-fdrv}(b) are well described by the scaling
form
\begin{equation}
        P(E, L) \sim L^{-\beta}~g( E / L^{\mu} ),
\label{e-fdrv}
\end{equation}
where $g$ is a scaling function which is constant for small arguments
and decays as a power law for large values of the argument.  As can be
seen in Fig.~\ref{f-fdrv}(c), a good data collapse is obtained with
the exponents $\beta = \mu = 1.00 \pm 0.05$ for $L \le 640$.  On the
other hand, for the two larger system sizes we again observe data
collapse but to a shifted curve.  The reason for this has to do with
the possibility of an avalanche running beyond the time at which
profiles are measured or a new grain is added:  for small $L$
almost every avalanche runs its course before a new grain is deposited
or the profile is measured, but for large $L$ we can still have running
avalanches before a new picture of the profile is taken 
implying that large
avalanches are not sampled and the entire distribution is shifted upwards
for large energies.
This hypothesis is confirmed by the fact
that for the largest values of $L$ the size of the biggest avalanche
does not seem to grow.

Another observation is that for the inaccurate estimation of the
energy, the value of the exponent seems to change from $1.53$ to
about $1.4$ [Fig.~\ref{f-fdrv}(b)]. 
The reason for this can be understood as a simple
finite-size effect and that it is only for a small region where
the true power-law behavior is observed. 
One way to circumvent the finite-size
effect is to measure instead the slope of the envelope which
is close to 1.53.
Thus, we expect the correct value of the exponent $\alpha$ to be $1.53$.

\begin{figure}[htb]
\centerline{
    \epsfysize=0.85\columnwidth{\rotate[r]{\epsfbox{\figdir/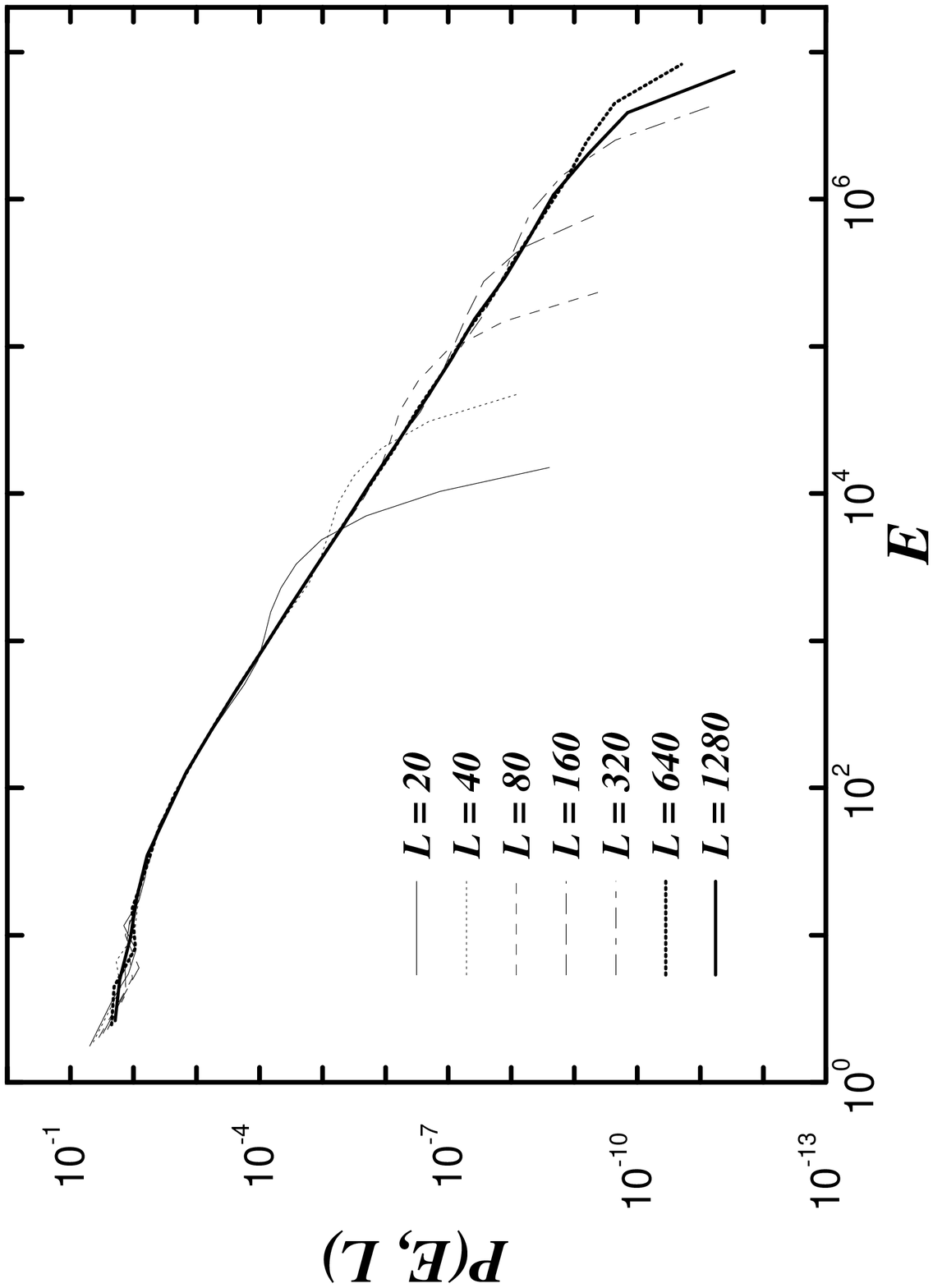}}}
}
\vspace*{0.5cm}
\centerline{
    \epsfysize=0.85\columnwidth{\rotate[r]{\epsfbox{\figdir/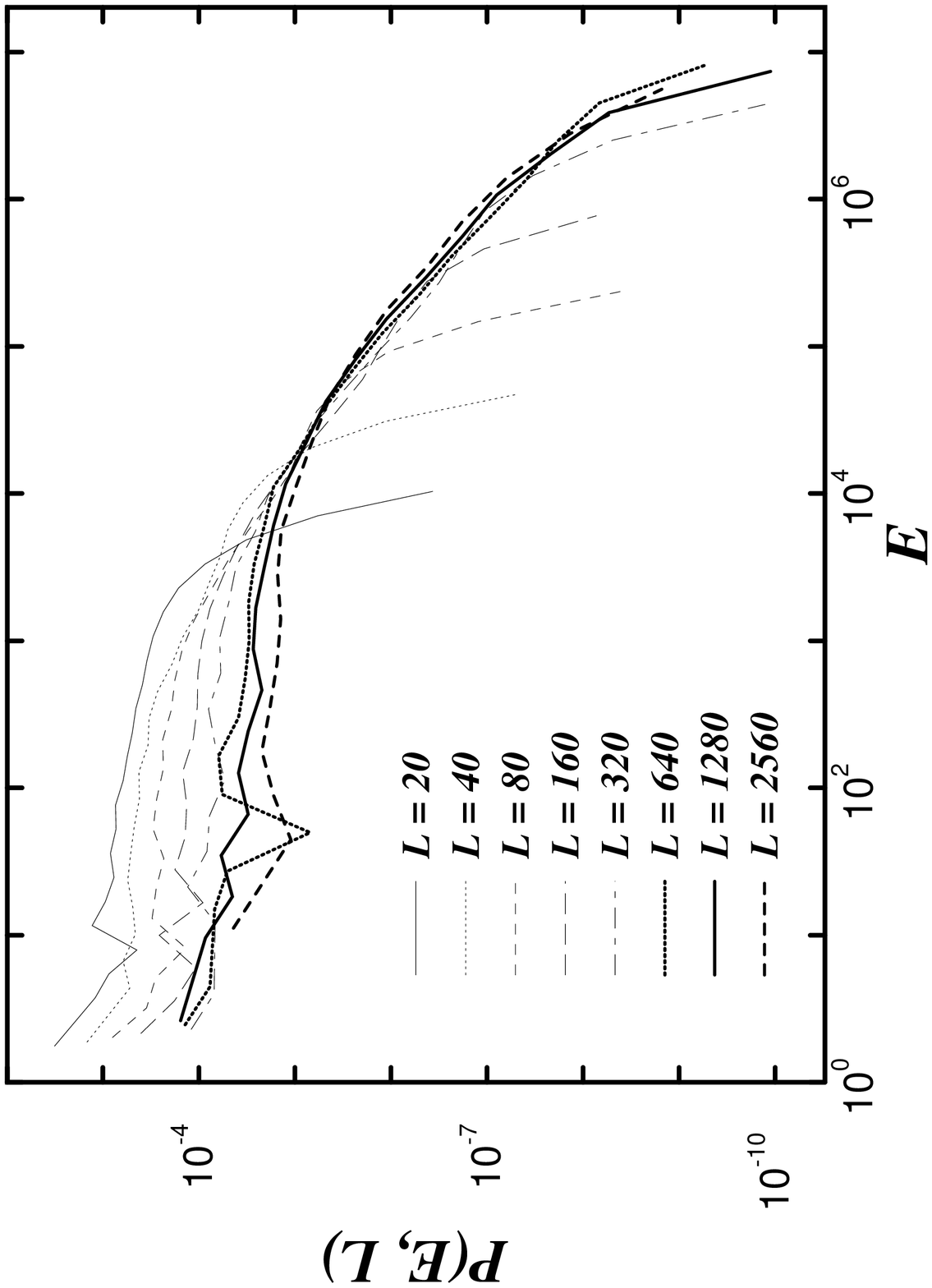}}}
}
\vspace*{0.5cm}
\end{figure}


\section{Conclusions}

We investigate energy avalanches in a one-dimensional rice-pile model.
We find that Eq.~(\ref{e-slow}) provides a good description of our
numerical results for the model when we drive it slowly.
Furthermore, when we drive the model in a way close to the
experimental conditions the numerical results are described by
Eq.~(\ref{e-fdrv}) in nice agreement with the experimental results in
\cite{frette-etal:1996}.  However, the numerical value for the
exponent describing the power law decay is 1.53, whereas the experiment
gave the value $\alpha_{\rm exp} \simeq 2$. Despite the fact that the
overall scaling form can be understood by the rice model studied here,
the disagreement in the values for $\alpha$ shows that the model needs
extensions.  One such possibility would be to include the effect of
the kinetic energy of the particles in addition to the friction
parameter $p$ considered here.

\section*{Acknowledgements}

We thank K. Christensen for discussions concerning
Ref.~\cite{frette-etal:1996}.  We acknowledge discussions with M. H.
Jensen, J. Krug, M. Mar\-ko\-sova, and K. Sneppen.  K.~B.~L. thanks
the Danish Natural Science Research Council for financial support.


\begin{thebibliography}{99}

\bibitem{BTW} 
P. Bak, C. Tang, and K. Wiesenfeld,
{Phys. Rev. Lett.} {59} (1987) 381.

\bibitem{KNWZ} 
L. Kadanoff, S. R. Nagel, L. Wu, and S.-m.\ Zhou, 
{Phys. Rev. A} {39} (1989) 6524.

\bibitem{CL}
J. M. Carlson and J. S. Langer, 
{Phys. Rev. Lett.} {62} (1989) 2632.

\bibitem{FF}
H. J. S. Feder and J. Feder,
{Phys. Rev. Lett.} {66} (1991) 2669.

\bibitem{Toner}
J. Toner, {Phys. Rev. Lett.} {66} (1991) 679.

\bibitem{Frette}
V. Frette, {Phys. Rev. Lett.} {70} (1993) 2762.

\bibitem{HK}
T. Hwa and M. Kardar, {Phys. Rev. A} {45} (1992) 7002.

\bibitem{luebeck}
S. L\"ubeck and K.~D. Usadel, Fractals {1} (1993) 1030.

\bibitem{JLN}
H. M. Jaeger, C.-h.\ Liu, and S. R. Nagel,
{Phys. Rev. Lett.} {62} (1989) 40.

\bibitem{H}
G.~A. Held, D.~H. Solina, D.~T. Keane, W.~J. Haag, 
P.~M. Horn, and G.~Grinstein, Phys. Rev. Lett. {65} (1990) 1120.

\bibitem{RVK}
J. Rosendahl, M. Veki\'c, and J. Kelley,
{Phys. Rev. E} {47} (1993) 1401.

\bibitem{RVR}
J. Rosendahl, M. Veki\'c, and J. E. Rutledge, 
{Phys. Rev. Lett.} {73} (1994) 537.

\bibitem{Feder}
J. Feder, Fractals {3} (1995) 431.

\bibitem{Hernan}
H.~A. Makse, S. Havlin, P. King, and H.~E. Stanley, preprint (1996).

\bibitem{frette-etal:1996}
V. Frette, K. Christensen, A. Malthe-S{\o}renssen, J. Feder, 
T. J{\o}ssang, and P.~Meakin, {Nature} {379} (1996) 49.

\bibitem{amaral-lauritsen:1996a}
        L. A. N. Amaral and K. B. Lauritsen, preprint (1996).
\bibitem{christensen-etal:1996}
        K. Christensen, A. Corral, V. Frette, J. Feder, and T. J\o ssang,
        preprint (1996).
\bibitem{markosova-etal:1996}
        M. Markosova, K. Sneppen, and M. H. Jensen, unpublished (1996).
\bibitem{amaral-lauritsen:1996b}
        L. A. N. Amaral and K. B. Lauritsen, preprint (1996).

\bibitem{maya}
	M. Paczuski and S. Boettcher, cond-mat/9603120.

\end{thebibliography}
\end{document}